\documentclass[apj]{emulateapj}
\usepackage{mathptmx}
\newcommand  \kms      {\ifmmode {\rm km\,s}^{-1} \else km\,s$^{-1}$\fi}

\newcommand  \ergs     {\ifmmode {\rm ergs\,s}^{-1} \else ergs s$^{-1}$\fi}
\newcommand  \ergcms   {\ifmmode {\rm ergs\,cm}^{-2}\,{\rm s}^{-1}
                        \else ergs\,cm$^{-2}$\,s$^{-1}$\fi}
\newcommand  \ergcmsA {\ifmmode{\rm ergs\,cm}^{-2}\,{\rm s}^{-1}\,{\rm\AA}^{-1}
                        \else ergs\,cm$^{-2}$\,s$^{-1}$\,\AA$^{-1}$\fi}
\newcommand \ergcmsHz {\ifmmode{\rm ergs\,cm}^{-2}\,{\rm s}^{-1}\,{\rm Hz}^{-1}
                        \else ergs\,cm$^{-2}$\,s$^{-1}$\,Hz$^{-1}$\fi}
\newcommand  \phcms    {\ifmmode {\rm ph\,cm}^{-2}\,{\rm s}^{-1}
                        \else ,ph\,cm$^{-2}$\,s$^{-1}$\fi}
\newcommand  \phcmsA   {\ifmmode {\rm ph\,cm}^{-2}\,{\rm s}^{-1}\,{\rm\AA}^{-1}
                        \else ph\,cm$^{-2}$\,s$^{-1}$\,\AA$^{-1}$\fi}
\def\micron{\ifmmode \mu{\rm m} \else $\mu$m\fi}
\def\kms{\ifmmode {\rm km\,s}^{-1} \else km\,s$^{-1}$\fi}
\def\Hubble{\ifmmode {\rm km\,s}^{-1}\,{\rm Mpc}^{-1}
        \else km\,s$^{-1}$\,Mpc$^{-1}$\fi}
\def\ergsec{\ifmmode {\rm ergs\;s}^{-1} \else ergs s$^{-1}$\fi}
\def\ergscm{\ifmmode {\rm ergs\,s}^{-1}\,{\rm cm}^{-2}
          \else ergs\,s$^{-1}$\,cm$^{-2}$\fi}
\def\ergscmA{\ifmmode {\rm ergs\,s}^{-1}\,{\rm cm}^{-2}\,{\rm \AA}^{-1}
          \else ergs\,s$^{-1}$\,cm$^{-2}$\,\AA$^{-1}$\fi}
\def\ergscmHz{\ifmmode {\rm ergs\,s}^{-1}\,{\rm cm}^{-2}\,{\rm Hz}^{-1}
          \else ergs\,s$^{-1}$\,cm$^{-2}$\,Hz$^{-1}$\fi}
\def\Msun{\ifmmode M_{\odot} \else $M_{\odot}$\fi}
\def\Lsun{\ifmmode L_{\odot} \else $L_{\odot}$\fi}
\def\qo{\ifmmode q_{0} \else $q_{0}$\fi}
\def\Ho{\ifmmode H_{0} \else $H_{0}$\fi}
\def\ho{\ifmmode h_{0} \else $h_{0}$\fi}
\def\qo{\ifmmode q_{0} \else $q_{0}$\fi}
\def\ao{\ifmmode a_{0} \else $a_{0}$\fi}
\def\to{\ifmmode t_{0} \else $t_{0}$\fi}
\def\Halpha{\ifmmode {\rm H}\alpha \else H$\alpha$\fi}
\def\Hbeta{\ifmmode {\rm H}\beta \else H$\beta$\fi}
\def\hb{\ifmmode {\rm H}\beta \else H$\beta$\fi}
\def\Hgamma{\ifmmode {\rm H}\gamma \else H$\gamma$\fi}
\def\Hdelta{\ifmmode {\rm H}\delta \else H$\delta$\fi}
\def\Lya{\ifmmode {\rm Ly}\alpha \else Ly$\alpha$\fi}
\def\Lyb{\ifmmode {\rm Ly}\beta \else Ly$\beta$\fi}
\def\hi{\ifmmode \mbox{{\rm H}\,{\sc i}} \else H\,{\sc i}\fi}

\def\ciii{\ifmmode {\rm C}\,{\sc iii} \else C\,{\sc iii}\fi}

\def\o5007{[O\,{\sc iii}]\,$\lambda5007$}

\def\neviIR {[Ne\,{\sc vi}]\,$7.6 \mu$m}
\def\pahIR {PAH\,$7.7 \mu$m}

\def\ne212m {[Ne\,{\sc ii}]\,$12.8 \mu m$}

\def  \kms         {\hbox{km s$^{-1}$}}          
\def  \ergs        {\hbox{erg s$^{-1}$}}              

\def  \La          {\ifmmode {\rm Ly}\alpha \else Ly$\alpha$\fi}
\def  \Ka          {\ifmmode {\rm K}\alpha \else K$\alpha$\fi}
\def  \Lb          {\ifmmode {\rm L}\beta \else L$\beta$\fi}
\def  \Ha          {\ifmmode {\rm H}\alpha \else H$\alpha$\fi}
\def  \Hb          {\ifmmode {\rm H}\beta \else H$\beta$\fi}
\def  \Pa          {\ifmmode {\rm P}\alpha \else P$\alpha$\fi}
\def  \CIIIb       {\ifmmode {\rm C}\,{\sc iii]}\,\lambda1909
                     \else C\,{\sc iii]}\,$\lambda1909$\fi}
\def  \CIV         {\ifmmode {\rm C}\,{\sc iv}\,\lambda1549
                     \else C\,{\sc iv}\,$\lambda1549$\fi}
\def  \MgII         {\ifmmode {\rm Mg}\,{\sc ii}\,\lambda2798
                     \else Mg\,{\sc ii}\,$\lambda2798$\fi}
\def  \OVI         {\ifmmode {\rm O}\,{\sc vi}\,\lambda1035
x
                     \else O\,{\sc vi}\,$\lambda1035$\fi}

\def \spitzer      {{\it Spitzer}}

\shorttitle{SEDs of QUEST Palomar-Green QSOs}
\shortauthors{Netzer et al.}
\slugcomment{Accepted by ApJ \today}

\begin{document}

\title{\spitzer\ Quasar and ULIRG Evolution Study (QUEST):\\
 II. The Spectral Energy Distributions of Palomar-Green Quasars }

\author{
Hagai Netzer\altaffilmark{1}
Dieter Lutz,\altaffilmark{2}
Mario Schweitzer,\altaffilmark{2}
Alessandra Contursi,\altaffilmark{2}
Eckhard Sturm,\altaffilmark{2}
Linda J. Tacconi,\altaffilmark{2}
Sylvain Veilleux,\altaffilmark{3}
D.-C. Kim,\altaffilmark{3}
David Rupke,\altaffilmark{3}
Andrew J. Baker,\altaffilmark{4}
Kalliopi Dasyra,\altaffilmark{5}
Joseph Mazzarella,\altaffilmark{5}
Steven Lord\altaffilmark{6}
}

\altaffiltext{1}
{School of Physics and Astronomy and the Wise Observatory, 
The Raymond and Beverly Sackler Faculty of Exact Sciences, 
Tel-Aviv University, Tel-Aviv 69978, Israel}
\altaffiltext{2}
{Max-Planck-Institut f\"ur extraterrestrische Physik, Postfach 1312, 
85741 Garching, Germany}
\altaffiltext{3}
{Department of Astronomy, University of Maryland, College Park, 
MD 20742-2421, USA}
\altaffiltext{4}
{Department of Physics 
and Astronomy; Rutgers, the State University of New Jersey; 136 
Frelinghuysen Road; Piscataway, NJ 08854-8019.}
\altaffiltext{5}
{Spitzer Science Center, 1200 E California Blvd, Pasadena CA 91125}
\altaffiltext{6}
{NASA Herschel Science Center, MC 100-22, Caltech, Pasadena CA 91125}

\begin{abstract}
This is the second paper studying  
the QSOs in the \spitzer\ QUEST sample. Previously  we presented 
new PAH measurements and argued that most of 
the observed far infrared (FIR) radiation is due to star-forming activity.
Here we present spectral energy distributions (SEDs) by supplementing our data 
with optical, NIR and FIR observations.  We define two sub-groups of 
``weak FIR'' and ``strong FIR'' QSOs,
and a third group of FIR non-detections. 
Assuming  a starburst origin for the FIR, we obtain   
``intrinsic'' AGN SEDs by subtracting a starburst 
template from the mean SEDs. The resulting SEDs are remarkably similar for all 
groups. They show three distinct peaks 
corresponding to two silicate emission features and a 3$\mu$m bump 
that we interpret as the signature of the hottest AGN 
dust.  They also display drops beyond 
$\sim 20\mu$m that we interpret as the signature of the minimum 
temperature ($\sim 200$~K) dust.
This component must be optically thin to explain the silicate emission and
the slope of the long wavelength continuum. We discuss the merits of 
an alternative model where most of the FIR emission is due to AGN heating.
Such models are unlikely to explain the
properties of our QSOs but they cannot be ruled out
for more luminous objects.
 We also find correlations between the luminosity at 5100\AA\ and two
infrared starburst indicators: L(60$\mu$m) and 
L(PAH 7.7$\mu$m). The correlation of L(5100\AA) with L(60$\mu$m) can be used to measure 
the relative growth rates and lifetimes of the black hole and the new stars.

\end{abstract}


\section{Introduction}
The spectral energy distributions, (SEDs), of active galactic nuclei (AGNs)
have been studied, extensively, over various energy bands and for different
sub-groups of the AGN population
(e.g. Sanders et al. 1989; Elvis et al. 1994, hereafter E94;
Scott et al 2004; Haas et al. 2003; Glikman et a. 2006; 
Richards et al. 2006, [hereafter R06]; Trammell et al 2006).
Such studies are essential for estimating the bolometric luminosities of AGNs,
for distinguishing between the various sub-classes, and for deriving the 
relationships between black hole (BH) mass, luminosity and accretion rate.
They are also required for understanding the details of the
energy production mechanisms at X-ray, UV, optical, infrared (IR) and 
radio energies. The shape of the observed SED depends on the 
BH  mass and accretion rate, the structure and inclination of the 
central accretion disk, the presence and the geometry of a dusty central structure,
the line of sight absorption and extinction, the presence of luminous starburst regions
in the host galaxy and the properties of 
any central radio source.

The various SED bands can be classified according to the origin of 
the emitted energy. Here we define ``primary radiation'' as the part 
produced within the inner 1000 gravitational radii (to include the entire
central accretion disk).
``Secondary radiation'' is the energy emitted outside of this radius,
due to  absorption and reprocessing of  primary radiation. The term ``intrinsic radiation''
will be used to define everything associated with the AGN, i.e. the primary
radiation as well as radiation from the vicinity of the BH 
due to the reprocessing of the primary radiation.

Dust in the immediate vicinity of the central source is known to be 
a major source of secondary radiation and is probably responsible 
for most of the emergent near-infrared (NIR) and mid-infrared (MIR) continuum.
How much intrinsic AGN-heated dust emission contributes to the far-infrared (FIR) 
emission spectrum has been an open question for years. This has been discussed
extensively in our first paper (Schweitzer et al. 2006, hereafter paper I)
where many relevant references are listed, and it is also a major topic of 
the present work. 
This is directly related to the possible contribution of star forming (SF) regions
to the FIR emission and the maximum and minimum  temperatures of the AGN-heated dust.

Most previous studies of AGN SEDs in the MIR, especially those focusing 
on the more luminous AGNs,  were limited by the sensitivity and 
resolution of earlier MIR instrumentation. This is now changing as 
new \spitzer\ observations are capable of providing high quality data 
on larger AGN samples. For example, \citet{lacy04} found that 
detection methods based on MIR colors are efficient tools for selecting 
all types of AGNs (and also starburst galaxies). The addition of optical colors makes such methods even 
more efficient (R06). Combinations of optical-MIR colors have been 
used by \citet{hatziminaoglou05} to derive mean type-I SEDs and to 
investigate the distribution of the bolometric luminosity in small 
AGN samples. R06 used a much larger sample of 259 Sloan Digital Sky 
Survey (SDSS) sources, and an optical-MIR color combination, to illustrate 
the diversity of AGN SEDs and to study the bolometric luminosity and 
accretion rates in such sources.

The present work is a continuation of our study of the QUEST sample 
which is described in detail in paper I. In short, 
we are studying QSOs, ultraluminous infrared
galaxies (ULIRGs), and the possible evolutionary connection between
the two using \spitzer-IRS.  The QSO
sample is largely drawn from that of Guyon (2002) and Guyon
et al. (2006). It consists of Palomar-Green (PG) QSOs
and covers the full ranges of bolometric luminosity ($10^{11.5-13}$ \Lsun),
radio loudness, and infrared
excess [$log (\nu L_{\nu} (60 \mu m)/L_{Bol} \sim  0.02-0.35$] spanned by the local
members of the PG QSO sample.
The ULIRGs in the sample  will be described, in detail, in
a forthcoming publication.

Unlike the SED studies mentioned above, we use detailed \spitzer\ spectroscopy
that allows us to expand on the earlier broad band work. A major goal 
of this study is to use MIR spectroscopy in combination with broad band 
FIR photometry in an attempt to investigate the origin of the FIR emission 
in intermediate luminosity QSOs. Such objects are classified
by R06 as ``normal type 1 quasars''. In paper I we focused on PAH 
features and the correlation of their luminosity with the FIR luminosity
of the sources. We detected clear PAH emission in 11 of our sources and 
argued for their likely presence in most other QSOs. We then used 
known relationships between PAH emission and star forming activity, 
and the great similarity of L(\pahIR)/L(60$\mu$m) in QUEST QSOs and 
the starburst dominated objects among the
QUEST ULIRGs, to argue that at least a third  and perhaps all of the 
50--100 $\mu$m luminosity in our QSO sample is due to star formation.
Here we present the entire 1--100$\mu$m SED of the QUEST QSOs and 
use the continuum properties to further investigate the PAH emission 
as well as other observed MIR features. A major goal is to
identify the ``pure'' or ``intrinsic'' AGN SED that is produced entirely 
by the central AGN with no starburst contamination.

Section 2 of the paper presents the MIR continuum spectra of the QSOs 
in our sample and shows derived SEDs for two sub-groups with weak and 
strong detected FIR emission, as well as for FIR non-detections. In \S3 we
introduce the intrinsic AGN-powered SED  and discuss the implications 
for AGN models and the AGN-starburst connection.

\section{The Infrared SED of QUEST QSOs}

\subsection{\spitzer\ spectra}
Our reduction and analysis procedure are 
described in paper I. The original QUEST sample includes 25 AGNs and four others, with
similar properties (I\,Zw\,1, PG\,1244+026, PG\,1448+273 and Mrk\,1014) were added
from other samples. The data presented here include two QSOs
(PG0844+349 and PG0923+201) that were not included in paper I because of  their later
observation dates. The combined sample now includes a total of 29 QSOs.
We use the flux calibrated 6--35 $\mu$m spectra and convert them to 
monochromatic luminosity 
($L_{\lambda}$ erg/sec/\AA) using a standard cosmology with
H$_0=70$ ${\rm km\,s^{-1}\, Mpc^{-1}}$, $\Omega_m=0.3$ and 
$\Omega_{\Lambda}=0.7$. Fig.~\ref{fig:all_spitzer_sed} shows the \spitzer\ 
spectra of these QSOs while the main spectral lines and features are
identified in Fig.~\ref{fig:pgaver}. The spectra are shifted in 
luminosity to avoid crowding and allow 
an easier inter-comparison of the overall spectral shape. They are also de-redshifted to compare
features in the same reference frame.  As clearly seen in 
this diagram, and in several of the diagrams shown in paper I,
there is a large variety of spectral shapes: some objects show a 
clear increase of $\lambda L_{\lambda}$ with $\lambda$ while in others 
there is a definite change of slope, and a clear decline for 
$\lambda > 15\mu$m. L(60$\mu$m)/L(15$\mu$m), a useful measure of the
FIR/MIR luminosity ratio, varies by up to a factor of 10 in our sample.
The luminosity range in our sample is a factor of about 100
in L(15$\mu$m) and  does not represent the entire type-I AGN
population since very high luminosity sources are missing from the
sample. Thus we are not in a position to investigate as large a range of SED
properties as done for example in R06. As noted in paper I, most
and perhaps all sources show signatures of silicate emissions peaking 
at around 10 and around 18$\mu$m. These features are the subject of a 
forthcoming paper.

\begin{figure*}
\plotone{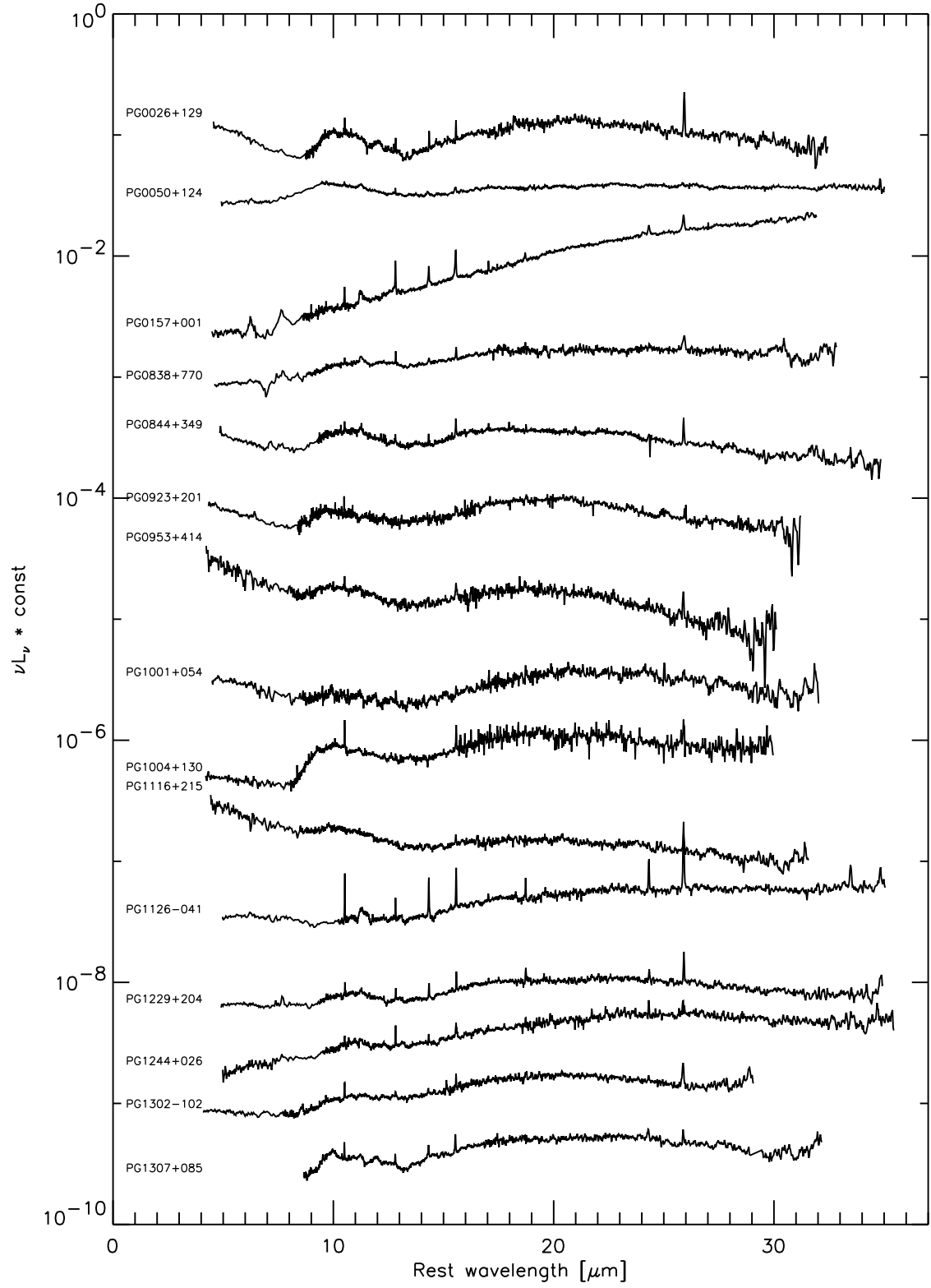}
\caption
{IRS spectra of  QUEST QSOs de-redshifted to rest-frame wavelengths and
given in $\nu L_{\nu} $ units. The spectra are shifted vertically
to allow clearer view. See Figure~\ref{fig:pgaver} for identification of
the main spectral features.}
\label{fig:all_spitzer_sed}
\end{figure*}

\addtocounter{figure}{-1}
\begin{figure*}
\plotone{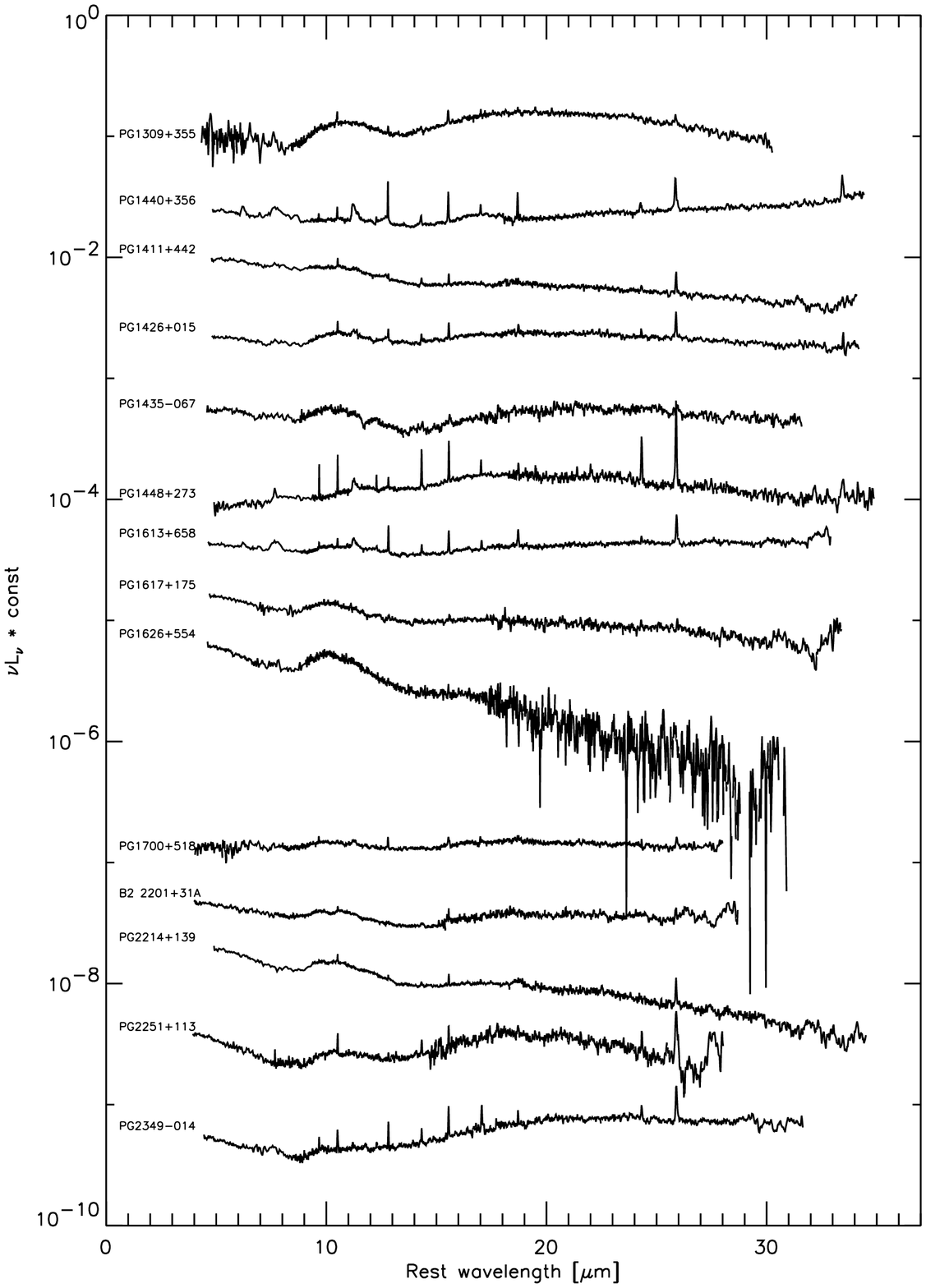}
\caption{continued}
\end{figure*}

\begin{figure}
\plotone{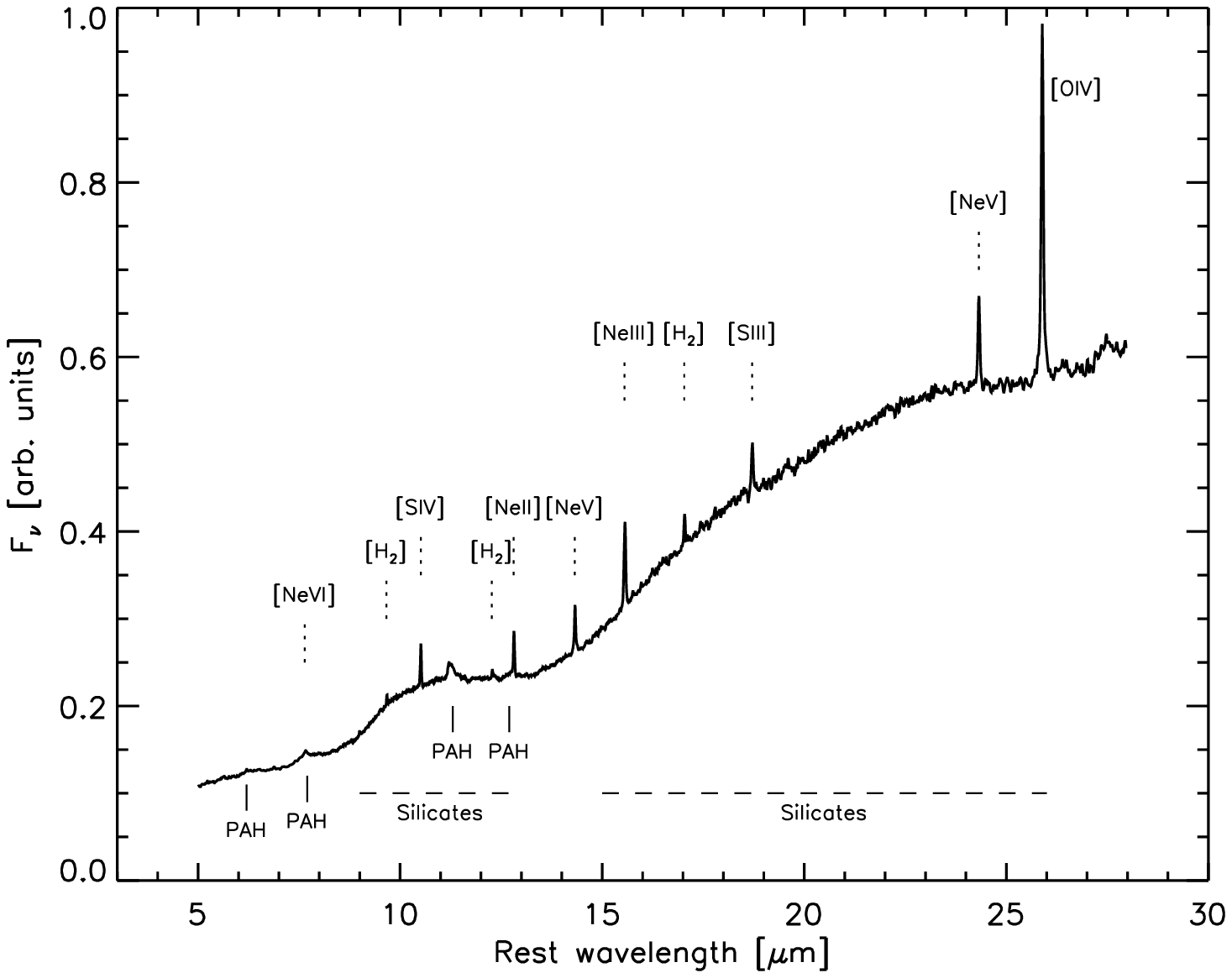}
\caption
{Average IRS spectrum of all QSOs except PG1307+085. The main emission lines
and features are marked.}
\label{fig:pgaver}
\end{figure}

In what follows we purposely avoid discussing sharp spectral features 
such as emission lines. These  will be discussed in a separate paper. 
All spectra shown 
below are presented on a rest wavelength scale 
to allow easier comparison among sources. For one of the targets 
shown in Fig.~\ref{fig:all_spitzer_sed} (PG1307+085), the low resolution 
IRS data down to 5$\mu$m are still 
proprietary to another program. This source is excluded from much of the
subsequent analysis.

\subsection{NIR and FIR data}
The basic \spitzer\ data and the FIR data obtained from the literature
for all but two recently observed sources are given in Paper I (table 1).
We have supplemented these with NIR data, and FIR data for the two new
sources, obtained 
from the literature.
The NIR data were obtained from various sources, in 
particular \citet{neugebauer87} and the 2MASS extended and point source
catalogs (Jarrett et al. 2000). Table~\ref{tab:targets} lists the adopted JHKL fluxes and 
references, where available. We have also added the optical luminosity, 
at 5100\AA, obtained from the original data of \citet{boroson92} 
and kindly supplied by T. Boroson. As discussed below, this makes an
important connection to the primary SED of the sources. Variability may
in principle affect the SEDs since  many of the optical/near-infrared 
data were taken about 20 years before the \spitzer\ observations. 
As for the new FIR data,  we followed the procedure in paper I and
 obtained the luminosity  at 60$\mu$m rest wavelength
by interpolating  the observed fluxes  at 
60 and 100$\mu$m. L(60$\mu$m) is very similar 
to the standard L(FIR) used extensively in the literature but relatively robust
to cirrus contamination at 100$\mu$m for these faint sources. The FIR 
luminosities computed this way for the two sources not included in paper I
are 10$^{10.38}$L$_\odot$ (PG0844+349) and $<10^{11.90}$L$_\odot$ (PG0923+201).
Hereafter we assume L(FIR)=L(60$\mu$m).

\begin{deluxetable*}{llrrrrrrc}
\tabletypesize{\footnotesize}
\tablewidth{0pt}
\tablecaption{QSO sample and supplementary data\label{tab:targets}}
\tablehead{
\colhead{Object}&
\colhead{z}&
\colhead{Log L$_{5100}$}&
\colhead{S$_{J}$}&
\colhead{S$_{H}$}&
\colhead{S$_{K}$}&
\colhead{S$_{L}$}&
\colhead{Refs}&
FIR class \\
\colhead{}&
\colhead{}&
\colhead{erg s$^{-1}$}&
\colhead{mJy}&
\colhead{mJy}&
\colhead{mJy}&
\colhead{mJy}&
\colhead{}&
\colhead{(for plot)}\\
\colhead{(1)}&
\colhead{(2)}&
\colhead{(3)}&
\colhead{(4)}&
\colhead{(5)}&
\colhead{(6)}&
\colhead{(7)}&
\colhead{(8)}&
\colhead{(9)}
}
\startdata
PG0026+129          &0.1420&44.66& 4.47& 5.82& 8.51&17.99&   N87&undetected\\
PG0050+124 (IZw1)   &0.0611&44.30&21.30&34.40&55.70&127.00&2M,S89&strong\\
PG0157+001 (Mrk1014)&0.1630&44.67& 6.06& 7.99&12.70&17.38&2M,N87&strong\\
PG0838+770          &0.1310&44.16& 2.29& 3.09& 4.79& 6.31&   N87&weak\\
PG0844+349          &0.0640&44.00& 6.03& 8.51&13.18&23.99&   N87&weak\\
PG0923+201          &0.1900&44.89& 2.90& 4.52& 8.79&14.70&2M,S89&undetected\\
PG0953+414          &0.2341&45.11& 3.39& 4.27& 7.76&15.49&   N87&undetected\\
PG1001+054          &0.1605&44.25& 1.47& 2.21& 4.10&     &   H82&strong\\
PG1004+130          &0.2400&45.23& 4.17& 4.27& 5.82& 9.12&   N87&weak\\
PG1116+215          &0.1765&45.13& 5.83& 8.64&16.15&32.50&2M,S89&undetected\\
PG1126-041 (Mrk1298)&0.0600&43.82&11.20&16.90&25.00&24.70&2M,S89&strong\\
PG1229+204 (Mrk771) &0.0630&44.13& 6.03& 8.51&13.18&23.99&   N87&weak\\
PG1244+026          &0.0482&43.26& 2.84& 3.66& 4.79&     &    2M&strong\\
PG1302-102          &0.2784&45.17& 3.36& 3.77& 4.89&     &   H82&strong\\
PG1307+085          &0.1550&44.87& 3.55& 4.32& 6.92&10.32&   N87&weak\\
PG1309+355          &0.1840&44.81& 3.33& 3.63& 5.76&     &    2M&undetected\\
PG1411+442          &0.0896&44.31& 5.62& 8.32&17.38&38.91&   N87&weak\\
PG1426+015          &0.0865&44.44& 5.89& 8.71&15.85&23.44&   N87&weak\\
PG1435-067          &0.1260&44.39& 2.52& 3.23& 5.84&     &    2M&strong\\
PG1440+356 (Mrk478) &0.0791&44.22& 9.77&15.14&25.12&38.91&   N87&strong\\
PG1448+273          &0.0650&43.99& 6.17& 8.41&11.89&15.14&   N87&weak\\
PG1613+658 (Mrk876) &0.1290&44.70& 4.57& 6.31&10.72&17.38&   N87&strong\\
PG1617+175          &0.1124&44.29& 4.27& 6.46&12.30&19.50&   N87&undetected\\
PG1626+554          &0.1330&44.44& 2.75& 3.43& 5.68&     &    2M&???\\
PG1700+518          &0.2920&45.68& 4.37& 6.53&12.74&30.90&   N87&weak\\
B2 2201+31A         &0.2950&45.91& 2.61& 4.24& 7.73&     &    2M&undetected\\
PG2214+139 (Mrk304) &0.0658&44.40&12.20&16.30&22.90&26.30&2M,M83&???\\
PG2251+113          &0.3255&45.63& 3.39& 4.68& 7.24&10.72&   N87&undetected\\
PG2349-014          &0.1740&45.21& 4.17& 5.57& 9.55&17.78&   N87&strong\\
\enddata
\tablecomments{\\
Col. (1) --- Source name.\\
Col. (2) --- Redshift.\\
Cols. (3) --- Continuum luminosity $\lambda$L$_\lambda$ at 
5100\AA\ rest wavelength (from spectra by T. Boroson, priv. comm.).\\
Cols. (4-8) --- near-infrared fluxes in the JHKL observed bands, and related 
references:\\
 2M -- 2MASS magnitudes, extended source catalog (Jarrett et al. 2000)
K20 isophotal magnitudes for
slightly extended sources, point source catalog magnitudes otherwise;\\
H82 -- \citet{hyland82};\\
M83 -- \citet{mcalary83};\\ 
N87 -- \citet{neugebauer87};\\
S89 -- \citet{sanders89}.\\
Col. (9) --- Classification as far-infrared strong/weak (relative to the 
mid-infrared), or undetected.
See text for definition and treatment of PG1626+554 and PG2214+139
}
\end{deluxetable*}

\begin{figure}
\plotone{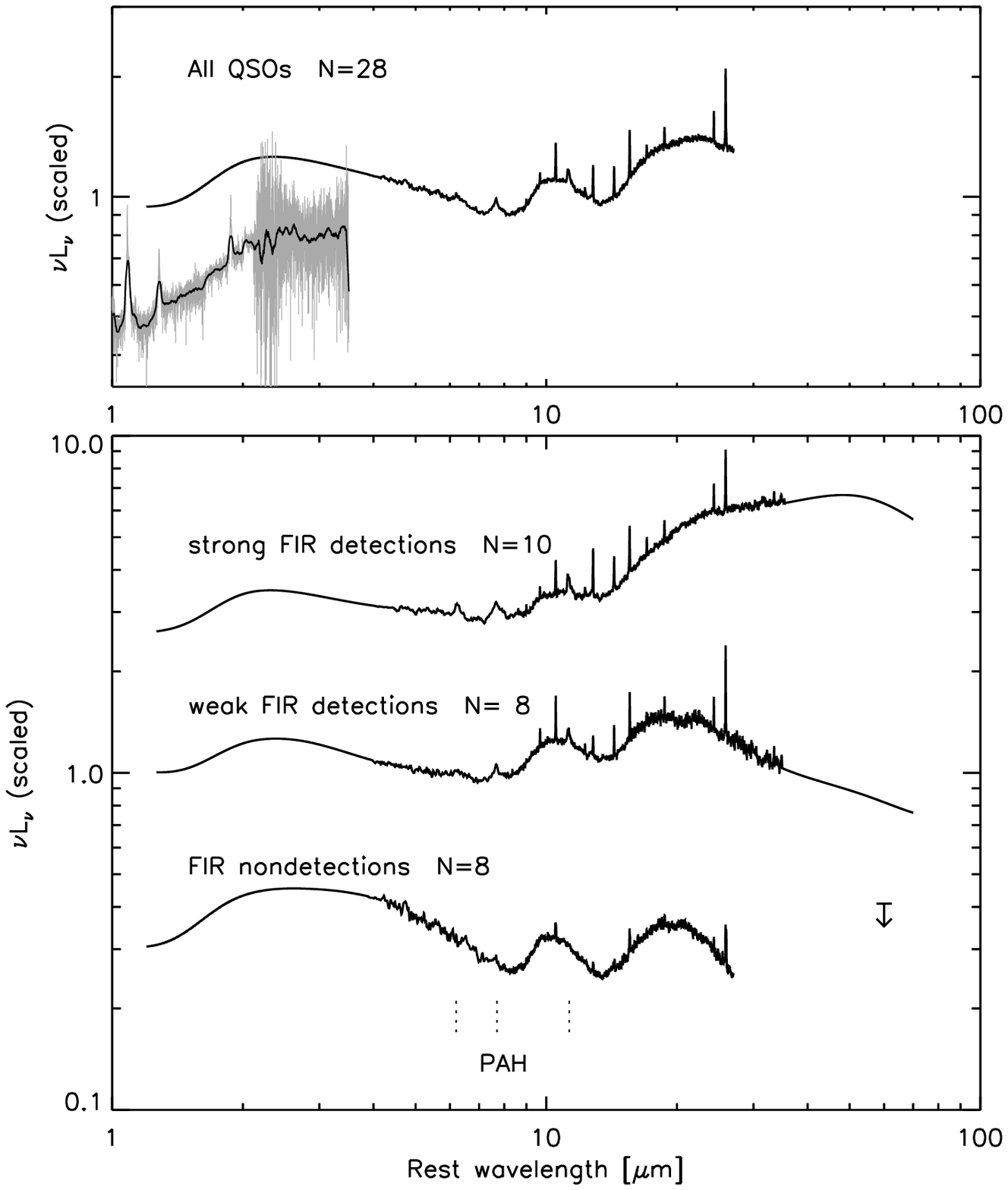}
\caption
{Top panel: Average spectral energy distribution for QUEST QSOs in
our sample, normalized at 6$\mu$m. For comparison, we overplot the
mean SED derived from ground-based
data at shorter wavelengths for a QSO sample of similar luminosity by
\citet{glikman06}, shifted by an arbitrary amount.
Bottom panel: Average SEDs for the three subgroups of
strong FIR detections, weak FIR detections, and FIR non-detections. Note the increase
of EW(\pahIR) from bottom to top.}
\label{fig:mean_SED_highres}
\end{figure}

\begin{figure*}
\epsscale{0.7}
\plotone{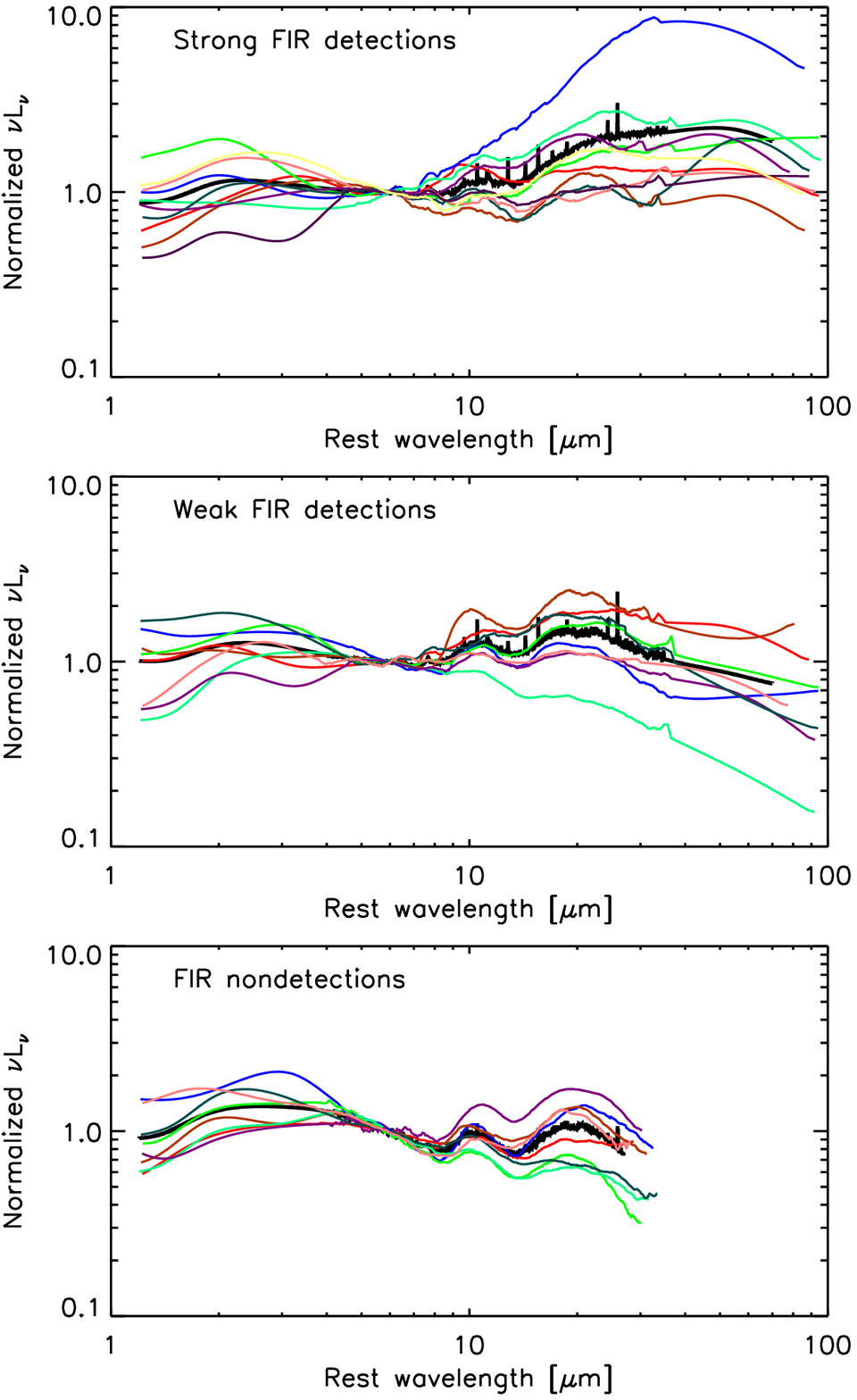}
\caption
{SEDs of 10 FIR-strong (top), 8 FIR-weak (middle) and 8
FIR-undetected (bottom) sources.
The central $\sim$5--35$\mu$m parts of the curves cover the range
of the \spitzer\ IRS spectra which are shown smoothed here. The short wavelength part
(optical, J, H, K and L photometry) and the long wavelength IRAS or ISO data
have been joined to the IRS spectra by interpolating splines.
The black heavy lines are the mean spectra of the two groups already
shown in Fig.~\ref{fig:mean_SED_highres}, the colored curves show the
scatter of individual objects around this mean. All spectra are normalized
at 6$\mu$m rest wavelength.
}
\label{fig:strong_weak_ir}
\end{figure*}

We extend the IRS spectra to shorter and longer wavelengths
by spline interpolations using the photometry obtained from the literature.
The dominant
sources of uncertainty in this procedure are the photometric uncertainties on the
data collected from the literature and the possible continuum variations at short wavelengths
since we are using non-simultaneous observations (see also Fig.~\ref{fig:strong_weak_ir}).
We also compared our data with the recent Glikman et al. (2006) composite obtained from a
sample of 27 AGNs that cover a redshift and luminosity range similar to the QUEST QSOs.
Their composite is of higher quality at shorter  wavelength and is shown next to our composite
SED in Fig.~\ref{fig:mean_SED_highres}.

\subsection{`Strong', `weak' and non-detected far-infrared sources}

To represent the range of far-infrared properties in
our sample, we divide the QSOs with FIR detections into
`strong FIR' and `weak FIR' emitters with a dividing line at 
rest L(60$\mu$m)/L(15$\mu$m)=1. Our definition thus implies `strong' or `weak' 
with respect to the mid-IR emission rather than in an absolute sense.
Out of the 21 QSOs in our sample with detectable FIR flux, two 
(PG2214+139 and PG1626+554) are problematic since the \spitzer\ spectra 
show a clear decline of $\lambda L_{\lambda}$ from 15 to
30$\mu$m yet the L(60$\mu$m) point lies clearly above the extrapolation of 
this spectrum. We suspect that the large aperture FIR data, close to the detection flux limit
of the IRAS and ISO instruments, are error-prone so we have chosen to
leave those sources out of the analysis. Also excluding PG1307+085, there remain 
 18 high quality SEDs extending to the FIR, out of 
which 10 are classified as strong-FIR sources and 8 
as weak-FIR sources. The remaining eight sources without
FIR detection define a third sub-sample; as explained in Paper I, these 
 are more likely to belong to the class of weak-FIR emitters 
and to also show very low equivalent width \pahIR\ emission, rather than having a rising IR
flux but a long wavelength flux below the detection limit.
We include the IRS and short wavelength data of those targets where
applicable.

Fig.~\ref{fig:mean_SED_highres} shows the mean SEDs of the full sample (excluding
PG1307+085) as well as of the above three FIR subgroups. The relative fluxes of the full SED are
given in Table~\ref{tab:seds}.
Setting aside emission lines and PAH features, all these 
mid-infrared SEDs show three pronounced peaks: one at 2--3$\mu$m 
plus the two silicate peaks at $\sim$10 and $\sim 18\mu$m. In addition to
these components, there is a differing amount of FIR emission
according to the specific subgroup. As discussed in paper I, FIR 
emission and PAH emission from the hosts of PG QSOs are correlated.
Indeed, Fig.~\ref{fig:mean_SED_highres} shows that the equivalent
width of \pahIR\ varies along with the far-infrared properties of the 
three subgroups.

The scatter of SED shapes within each group is indicated in Fig.~\ref{fig:strong_weak_ir}.
The SEDs clearly show
upturns and downturns at long ($\lambda > 40 \mu$m) wavelengths
justifying the terms strong and weak FIR sources. However, there is a 
range of properties within each group and a continuous transition from 
one to the other. We also show, in thick black lines,
the three mean SEDs. We prefer the use of mean rather than median spectra 
because they better preserve the integrity of features in the SEDs and 
because of the small number of sources. However, the overall differences between the 
median and the mean SEDs are not large.

The individual source SEDs sometimes show structure at $\lambda <5\mu$m
which may be due to uncertain (literature) photometry and/or variability.
However, the detections
of the three peaks noted above in all mean SEDs of Fig.~\ref{fig:mean_SED_highres} suggest
this result is robust.

The results shown here are consistent with the conclusions of paper I that 
most, and perhaps all QSOs show some level of PAH emission,  and that any 
classifications into PAH-detected and non-PAH-detected categories is likely  due to 
a combination of real differences in EW(\pahIR) (or other PAHs)
 and observational limitations (e.g. aperture affects).
It is therefore not surprising that both groups, of strong and weak 
FIR sources, include sources with detectable \pahIR. We expect future 
higher signal-to-noise spectra to reveal currently non-detected PAHs.
In summary, the observations of the QUEST QSOs suggest that we are 
sampling a distribution in all three properties: PAH luminosity, 
FIR luminosity and SED shape, with a tendency for larger FIR luminosity 
sources to have larger L(\pahIR) and a long wavelength upturn of their SED.

\subsection{Correlations between L(FIR) and L(PAH) and the primary AGN luminosity}
\label{sect:firopt}

So far we have only considered the NIR, MIR and FIR parts
of the SED. These are mostly due to thermal emission from cool and hot dust 
which, by definition,  are secondary sources of radiation.
Our data set also includes 4000--6000\AA\ spectra of all sources. 
This radiation is thought to be emitted by the central accretion disk and 
is thus  primary. We can directly compare this 
radiation with several of the IR components, including those suspected to
be of starburst origin. In the following we use L(5100) to 
specify $\lambda L_{\lambda}$ at 5100\AA. This quantity is
relatively easy to measure and is widely used to derive the BLR size and 
the central BH  hole masses in AGNs (e.g. Kaspi et al. 2005).

\begin{figure}
\plotone{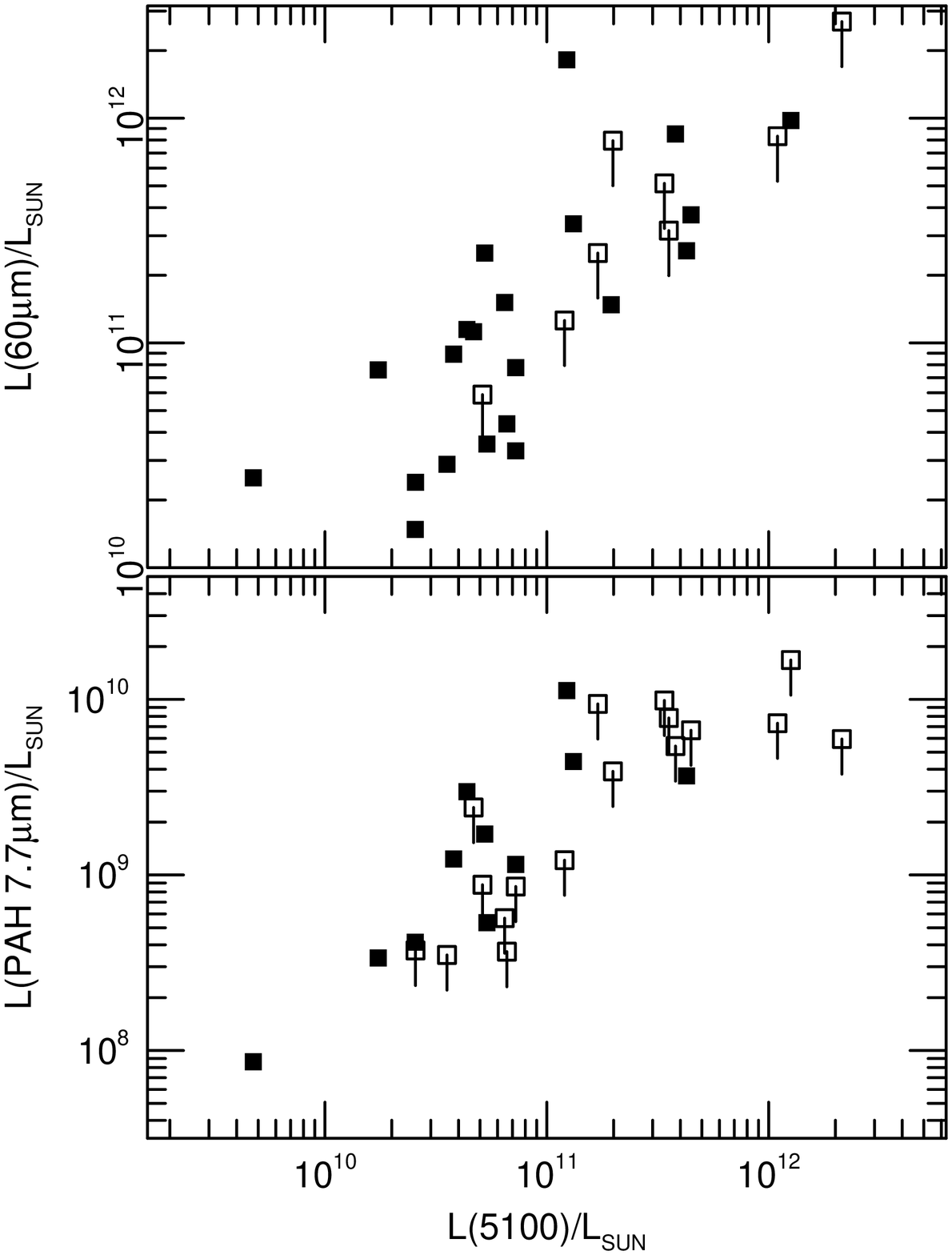}
\caption
{Top: The correlation of the optical (5100\AA) and FIR ($60\mu$m) continuum luminosities.
Bottom: L(5100) vs. L(\pahIR) showing detections (full squares) and upper
limits (empty squares).
}
\label{fig:L60_vs_L5100}
\end{figure}

The correlation of L(5100) with L(60$\mu$m) is shown in 
Fig.~\ref{fig:L60_vs_L5100} where we plot data for all the QUEST QSOs
(21 detections and 8 upper limits).
The diagram exhibits a very strong correlation over more than two orders
of magnitudes in L(5100) with a slope $\alpha \simeq 0.8$
($ L(60\mu {\rm m}) \propto L(5100)^{\alpha}$)
for the 60$\mu$m detected sources. The error on the slope is about 0.16, again using
only real detections. Assuming the upper limits represent real detections we get a very
significant correlation. 
However, at this stage the result is not very sound since testing for the correlation
of the observed fluxes (i.e. the observed $\lambda f_{\lambda}$ at 60$\mu$m rest wavelength
 vs. $\lambda f_{\lambda}$ at 5100\AA\ rest wavelength)
gives a much weaker dependence (rank correlation of 98\% singificance
when all sources are included). This is likely the result of the large spread in L(FIR) and
L(60$\mu$m)  together
with the very limited flux range of the sample. A larger sample is required  to verify this
relationship. Notwithstanding this limitation, we proceed by assuming that
the two luminosities are indeed correlated and discuss the implications to AGN physics. 

Estimates of the bolometric luminosity, $L_{\rm bol}$,  of unreddened AGNs 
(e.g. Shemmer et al. 2004; Marconi et al. 2004; Netzer \& Trakhtenbrot 2007)
are  5--10 times larger than L(5100).
In this context
$L_{\rm bol}$ applies to the primary continuum radiation and the above references suggest that
the bolometric correction
is probably luminosity dependent. For the luminosity range of the QUEST QSOs,
$L_{\rm bol} \sim 7$L(5100).
This issue is crucial to our work and we need to describe it in more detail.

The multi-wavelength multi-object study by R06 followed the  
Elvis et al. (1994) approach  and recommended
a large bolometric correction ($\sim 12$) relative to L(5100).
The number was obtained by 
integrating over the entire continuum, including the infrared part.
 According to our definition, there is a clear distinction between the primary  and secondary sources of radiation.
The formaer is the result of the accretion disk and its corona (or alternative X-ray producing mechanism) and
the latter is due to reprocessed radiation. The bolometric correction factor of $\sim 7$ used here assumes isotropic
radiation by the primary source at all wavelengths. Under this assumption, the R06 procedure involves
 double counting and hence the larger bolometric correction factor obtained by these authors.
 In our sample L(FIR) is, on the average, very similar to L(5100)
(see Fig.\ref{fig:L60_vs_L5100}). Given the above bolometric correction we find that for QUEST QSOs, L(FIR) is
roughly 10--20\% of the (primary) AGN bolometric luminosity.

A similar correlation to the one shown in Fig.\ref{fig:L60_vs_L5100} was 
also found by \citet{haas03} in their study of a large sample of PG QSOs 
(see Fig.~4 in their paper). That sample contains a small number of sources 
with optical luminosities that are significantly higher than those 
considered here . These sources seem to deviate from the almost 1:1
relationship found here with a hint that the FIR luminosity  levels off
at around $10^{13}$\Lsun. (Note that Haas et al. use L$_B$ which is somewhat different
than L(5100) used here).

We also test the correlation of L(5100) with L(\pahIR) which was not available in 
the \citet{haas03} sample. As noted earlier, there are 11 sources with 
direct L(\pahIR) measurements and 17 with upper limits. All are 
plotted in the lower part of Fig.~\ref{fig:L60_vs_L5100}.
The diagram suggests a tight correlation between the two properties with 
the more optically luminous sources also the ones with larger 
L(\pahIR). This was not investigated statistically since the
majority of sources do not have direct PAH detections. The strong correlation  is
not surprising given the correlations of both L(5100) and L(\pahIR) 
with L(FIR) and the fact that most measured upper limits on L(\pahIR) are 
likely to be within a factor of 2--3 of the real L(\pahIR) (paper I). 
While we are not in a position to test this relationship for sources 
with L(5100)$>10^{12}$\Lsun, we note the tendency of upper limits at the high-L end
of the diagram to fall below the relationship seen for the lower luminosities. This
is now confirmed by the recent work of Maiolino et al. (2007) who studied much higher
luminosity QSOs.

\section{Discussion}

\subsection{Possible origins of the FIR emission}
The present study extends the work of \citet{sanders89}, \citet{elvis94}, 
\citet{kuraszkiewicz03}, \citet{haas03}, R06 and others who studied the 
IR-SED of various subgroups of AGNs. Many of the spectra used to construct 
those SEDs are of sources that were either found in X-ray selected samples 
(e.g. the sources in \citet{elvis94} and \citet{kuraszkiewicz03}) or in UV
selected samples (the PG samples of \citet{sanders89} and \citet{haas03}).
Our QUEST sample is very similar to the  \citet{sanders89} sample and 
also to the other samples 
containing UV selected X-ray bright QSOs. Two important differences
are the small fraction of radio-loud sources in our QUEST sample (5 out 
of 29 QSOs compared with about 50\% of the sources in 
\citet{haas03}) and the smaller range in optical luminosity (only two  
orders of magnitude). Despite the relatively small luminosity 
range, the diversity in spectral properties in our sample is large
with a clear distinction between weak-FIR and strong-FIR sources 
(Fig.~\ref{fig:strong_weak_ir}). Most important regarding the comparison 
with earlier works is the greater level of detail over the 6--35$\mu$m 
range where \spitzer\ IRS is clearly superior to previous instruments. In 
this respect our SEDs are superior to the (much larger number of) broad 
band R06 SEDs. This improved resolution results in the detections of 
previously unobserved PAH features (paper I), in a real correlation between L(\pahIR) and
L(FIR) in QUEST QSOs,  and in a more detailed view
of the shape and the strength of several MIR features, such as the 
silicate bumps centered at around 10 and 18$\mu$m.

A major goal of the present investigation is to isolate the AGN-powered 
IR spectrum of the QUEST QSOs. This requires an understanding of the 
origin of the FIR emission and a comparison with theoretical
models that predict the expected dust emission under different conditions.
There are a number of such models in the literature including those of
Pier \& Krolik (1992; 1993),  \citet{granato94}, \citet{efstathiou95},
\citet{nenkova02} and \citet{kuraszkiewicz03}.
They address various possibilities regarding the dust distributions,
the orientation of the central torus, and the overall geometry.
In general, such models can be divided into two groups:
those assuming a continuous gas distribution (all models except those
of Nenkova et al.) and those assuming a clumpy dusty medium
(\citet{nenkova02}, see also \citet{elitzur04}).
Examples of fits to observed IR spectra are shown in
\citet[][models of the SEDs of three QSOs]{kuraszkiewicz03},
\citet[][see Fig.~5]{pier93} and \citet[][e.g. Fig.~3]{nenkova02}.
Key issues in such models are the agreement with the observed NIR-MIR spectrum and the
question of whether the FIR emission is due to the same central structure or whether it is 
produced, independently, by isolated kpc-scale star-forming regions in the host galaxy.

Kpc-scale luminous star forming regions have been proposed, in several
papers, as the origin of the FIR radiation in QSOs and in lower luminosity 
Seyfert galaxies (e.g. Rowan-Robinson 1995; Barthel  2006).
\citet{haas03}
address this possibility in their comprehensive investigation of a sample 
of 47 PG QSOs which covers about three orders of magnitude in optical 
continuum luminosity. In particular, they provide estimated dust temperatures that are required
to explain the observed FIR spectrum of all sources. These estimates are 
based on broad band measurements of the 20--100$\mu$m continuum and 
suggest a typical dust temperature of 30--50 K for most sources, similar to 
the range found for nearby ULIRGs by \citet{klaas01}. According to
\citet{haas03}, the high redshift highest luminosity sources in the sample 
exhibit a warmer IR continuum that peaks in the MIR part of the spectrum. 
In such sources, the entire 50--150$\mu$m part of the spectrum is interpreted as 
the Rayleigh-Jeans tail of an AGN-heated dust component. \citet{haas03} further 
searched for PAH emission in those sources. While they did not have any 
detections, they commented on the fact that the derived upper limits are
consistent with the assumption that the entire FIR luminosity is of 
starburst origin, except for the few most luminous QSOs in their sample.

Further general support for the likely SF origin of the FIR emission 
comes from the correlation of radio and infrared properties. According to 
\citet{haas03}, the observed L(1.4 GHz)/L(FIR) ratio in radio quiet QSOs
(about 50\% of the sources in their sample) is very similar to the ratio 
observed in starburst galaxies \citep{condon92}, where the FIR emission 
is due to starburst heated dust, and the radio emission due to supernovae
is proportional to the star formation rate. Radio-loud QSOs 
contain an additional,  more powerful radio source which is associated 
with the compact active core of the AGN. The \citet{condon92} relation applied
to our sample seems to provide a firm lower limit to the radio emission of all
PG quasars. This strengthens the assumption that powerful star formation is
taking place in most, and perhaps all such sources.

Given all this, we proceed to
discuss two different scenarios. The first involves two separate
IR sources: an inner structure that emits the 1--40$\mu$m continuum
and a surrounding  SF region that emits the FIR continuum and the associated
PAH emission features. The second requires only one large dusty structure
that produces the entire IR spectrum by absorption and re-emission of
the primary source radiation at different distances.
Along the way we also discuss caveats related to the expected range in L(FIR)/L(PAH).

\subsection{The intrinsic SED of type-I AGNs}

\subsubsection{Starburst produced FIR continuum}
We first consider the possibility that the infrared SED of the QUEST QSOs
contains two distinct components: one originating in a dusty central 
structure and the other in extended SF regions.
As shown in paper, the QUEST observations are consistent with the assumption that at least
one third, and perhaps almost all the FIR emission is due to starbursts.
Given 100\% starburst contribution we find that the starburst luminosity in our sample
is between $1.6 \times 10^{10}$ to $2.5\times 10^{12}$\Lsun\ and the
corresponding (somewhat model dependent) star formation rate
is between about 2 and 300 solar masses per year.
We caution that the 
results presented below may not apply outside the luminosity range 
considered here. Thus sources like  
those observed by \citet{haas03}, that are an order of magnitude more 
luminous than the most luminous QUEST QSOs, may contain a different
combination of warm AGN-heated and cold starburst-heated dust.

We proceed to produce a template starburst SED that will be subtracted
from the spectra of all the QUEST QSOs. The best estimate for such a
template is the mean spectrum of the 12 starburst dominated ULIRGs in our
QUEST sample that do not show AGN indicators like strong mid-IR continuum
or high excitation lines like [{Ne\,{\sc v}] or [{O\,{\sc iv}]  and also do not show
strong MIR absorption (the SB-ULIRG group also discussed in paper I \S4.1).
The detailed  properties of those 12 ULIRGs will be discussed in a future paper, here
we only use their composite spectrum.
This choice is preferable to the use of templates
based on observations of nearby starburst galaxies since these are not only
different in their starburst properties but are also subjected to aperture,
metallicity and luminosity effects.

Our starburst dominated ULIRG template was obtained by normalizing
the 12 spectra to have the same 60$\mu$m flux and then taking the mean at
every wavelength. The individual spectra show modest spread
around the mean for all $\lambda > 5 \mu$m, which is the part of the
spectrum relevant for our analysis. This starburst dominated ULIRG
template was subtracted from the two mean QSO spectra (weak and strong FIR
emitters), with a scaling consistent with the assumption that 
most of the 50--100 $\mu$m emission is due to star formation. The normalization that
was adopted made the long wavelength part of the residual spectrum,
after subtraction (i.e. the part assumed to be the intrinsic AGN SED), consistent 
with optically thin dust emission. The presence of such an optically thin 
AGN dust component is 
unambiguously indicated by the silicate emission features. We have used 
optically thin dust models that were computed for fitting the silicate emission
in QUEST QSOs (Schweitzer et al., in preparation). This method dictated the fractional amount
of starburst contribution at those wavelengths albeit with a rather large uncertainty.
It resulted in $\sim 87\%$ due to starbursts at 60$\mu$m  
for the  
strong FIR case and $\sim 80\%$ for the weak FIR case.
Fig.~\ref{fig:sedpgsub} shows the two mean QSO spectra before and after
this subtraction. The adopted starburst template is consistent with the 
L(\pahIR)-L(FIR)
correlation for QSOs (Fig. 4 of paper I) and the similarity of
\pahIR/FIR ratios in PG QSOs and star-bursting ULIRG stated there.
The scaling is also independently supported 
by the absence of PAH residuals in the subtracted spectra of 
Fig.~\ref{fig:sedpgsub}.
We thus obtain ``intrinsic AGN'' SEDs for each of the two groups.

\begin{figure*}
\epsscale{1.0}
\plotone{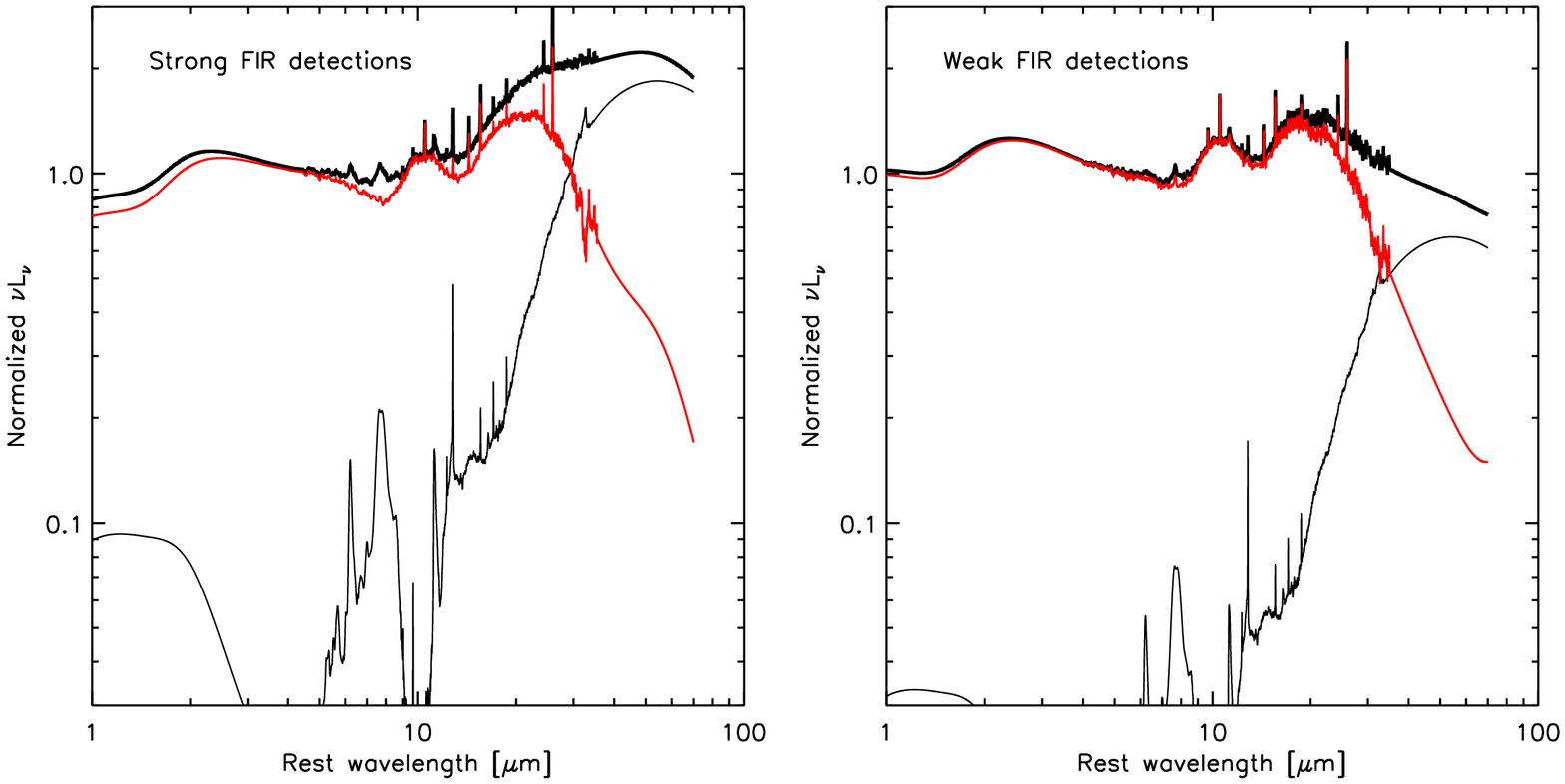}
\caption
{Normalized mean SEDs for strong FIR QSOs (left, top curve)  and weak FIR
QSOs (right, top curve).
The adjacent red SED curves show ``intrinsic'' AGN SEDs obtained by the
subtraction of the scaled mean starburst (ULIRG) spectrum (shown in black) from
the mean SEDs. (See text for explanation).
}
\label{fig:sedpgsub}
\end{figure*}

The assumption that most of the FIR emission is due to a starburst 
component results in a clear drop of the computed AGN SED at long wavelengths. 
While the exact $\lambda >30\mu$m slope of  depends
on the adopted scaling, a clear decay in $\lambda$F$_\lambda$ beyond 30$\mu$m
cannot be avoided without leaving significant PAH residues which would
indicate failure to subtract star formation. In contrast, our intrinsic
AGN SEDs show only some \neviIR\ emission left on top of the continuum in 
the 6--8$\mu$m region. As evident from Figs.~\ref{fig:all_spitzer_sed} and 
\ref{fig:sedpgsub}, almost all of our sources, and the two median SEDs,
contain noticeable silicate features that must originate in optically thin
regions of dust emission. The inferred long wavelength drop of our
intrinsic AGN SED is consistent with this component since it suggests a slope which
is steeper than a blackbody slope. A
flatter  SED (in particular for the FIR-strong sources) 
would require an alternative explanation.

Inspection of Fig.~\ref{fig:sedpgsub} reveals a high degree of similarity between
the intrinsic AGN SEDs of weak and strong FIR emitters. They both show a 
relatively flat spectrum in $\lambda L_{\lambda}$ with three distinct 
``bumps'': two corresponding to the silicate features at 10 and 18$\mu$m, 
and a third which is centered at around 3$\mu$m.
The average SED of the sources with FIR upper limits is also in agreement with
the two mean SEDs over the wavelength range where it is available, and also shows
the same three bumps. Table~\ref{tab:seds} lists the intrinsic 1--40 $\mu$m spectrum of the weak FIR group
in normalized flux units. 

\begin{deluxetable*}{ccc}
\tabletypesize{\footnotesize}
\tablewidth{0pt}
\tablecaption{Average QSO SEDs\label{tab:seds}}
\tablehead{
\colhead{Rest Wavelength}&
\colhead{Observed QSO SED}&
\colhead{Intrinsic AGN SED for FIR-weak QSOs}\\
\colhead{$\mu$m}&
\colhead{$\lambda$F$_\lambda$ (arb.units)}&
\colhead{$\lambda$F$_\lambda$ (arb.units)}
}
\startdata
 1.202&0.944&0.974\\
 1.216&0.944&0.973\\
 1.230&0.945&0.972\\
 1.245&0.946&0.972\\
 1.259&0.947&0.971\\
 1.274&0.948&0.971\\
 1.288&0.949&0.971\\
 1.303&0.951&0.971\\
 1.318&0.953&0.971\\
 1.334&0.956&0.972\\
 1.349&0.959&0.973\\
\enddata
\tablecomments{\\
Col. (1) --- Rest wavelength\\
Col. (2) --- Average observed SED of 28 PG QSOs
             (see Fig.~\ref{fig:mean_SED_highres} top, smoothed)\\
Col. (3) --- Average intrinsic AGN SED of 8 'FIR-weak' PG QSOs, after
            subtraction of a starburst component
             (see Fig.~\ref{fig:sedpgsub} right, smoothed)\\
Full SEDs are given in an electronic table.
}
\end{deluxetable*}

The short wavelength
feature has been noted in various earlier papers, most recently by
\citet[][see references to earlier work in this paper]{glikman06}
who could only observe the short wavelength side of the feature and 
modeled it as a combination of a nonthermal powerlaw and a
1260\,K blackbody. It extends from below
1$\mu$m (the ``1 micron inflection'' in R06) to about 5--8 $\mu$m and
is  better seen in our new SEDs because of the much clearer view of 
the 5--10 $\mu$m part of the spectra, where the long wavelength upturn 
of this feature is included in the \spitzer\ spectral range. We interpret 
this feature as the signature of the hottest dust in the AGN inner structure,
at a temperature of 1000--1500 K. Obviously, a realistic torus will radiate over
a range of temperatures. We also caution that there is a relatively 
wide range,  and a large scatter, among the individual sources of our sample 
in this part of the spectrum and some of the data (e.g. the L-band photometry)
are incomplete. Given this, the difference between the two intrinsic SEDs 
presented here is surprisingly small.

In summary, the SF-dominated scenario for the FIR implies similar AGN SEDs
for all sources,  showing 
three distinct NIR-MIR humps and very similar long wavelength 
($\lambda>20~\mu$m) slopes. Such a slope is consistent with the
observed silicate features and can be attributed to the  absence of large amounts
of AGN-heated dust with temperatures below about 200K.

\subsubsection{AGN-produced FIR continuum}

An alternative view is that direct AGN heating, followed by re-radiation 
of cool distant dust, is the origin of a large fraction of the observed FIR emission. Such 
scenarios have been discussed in several earlier papers including 
\citet{sanders89} and \citet{haas03}. Successful 
models of this type must explain the required range of dust temperatures 
as well as the FIR luminosity and the strength of the PAH emission.
In this case, the differences between strong-FIR and weak-FIR emitters reflect a
range in dust temperature. For sources of
similar primary luminosity, differing SEDs imply
different dimensions and/or geometry of the inner dusty structure.

For the QSOs under study, the FIR luminosity is comparable to L(5100) 
which, in turn, represents some 15\% of the bolometric luminosity 
(\S\ref{sect:firopt}) (the ratio has a  weak dependence on L(5100) that we ignore here).
 Thus $L_{12} \simeq 7 L_{12}(FIR)$,
where $L_{12}$ and $L_{12}(FIR)$ are the bolometric (primary) and
 FIR luminosities in units of $10^{12}$ \Lsun, respectively.
In general, continuous dust distribution models, such as those published by
\citet{pier93}, would fail to produce such a large FIR luminosity
since the assumed torus geometries result in a strong attenuation of the central
source radiation at large distances. A somewhat different geometry, of
a flaring or warped disk, can solve the FIR energy budget problem but
introduces a related problem of extremely large dimensions. This is easily
understood if we consider the case of a dusty region which
is fully exposed to the central source of radiation. In such gas
\begin{equation}
T_{dust} \simeq 1500 r_{pc}^{-t} L_{12}^{1/4} \, {\rm K} \,\, ,
\end{equation}
(e.g. Phinney 1989)
where $r_{pc}$ is the distance from the central source in pc and
$t=2/(4+s)$. For  optically
thick dust $s=0$ and for other cases it describes the frequency variation
of the dust cross section ($\sigma_{\lambda} \propto\lambda^{-s}$). For most
cases of interest $s=1-2$.

The present scenario suggests that the 50--100 $\mu$m emission of the QUEST QSOs
is mostly from dust with  $T_{dust} \simeq 40-65$\,K.
For a typical low luminosity source in our sample $L_{12} \simeq 0.2$, which 
for the optically thick case with $T_{dust}=50$\,K gives 
$r \simeq 0.4$ kpc. For optically thin dust with $s=1.5$ we get 
$r \simeq 1.7 $ kpc. The derived dimensions 
for the highest luminosity QUEST QSOs, with $L_{12} \sim 17$, are 
an order or magnitude or more larger.

Given the inferred dimensions, and the fraction of the bolometric luminosity
absorbed by the dust, we find that the distant dusty gas in the most 
luminous QUEST QSOs must rise to a height of at least 0.6 kpc, and
perhaps much higher, above the galactic plane. Such a huge structure must 
have unusual dynamical consequences and is not consistent with a stable, 
relaxed system. The limitations are probably less severe in interacting 
systems where warped galactic disks may be involved. The solid angle in 
this case can be considerably larger, especially during the later phases 
of the merger.

The clumpy models of \citet{nenkova02} come closer to the observed 
dimensions since, in those cases, nearby clumps can contribute to the IR 
emission from their hot (illuminated) as well as their cold (back-side)
faces. In such cases, the dimensions can be smaller and the only firm lower
limit on $r_{pc}$ is obtained from simple blackbody considerations.
For a partially filled radiating spherical surface of solid angle $\Omega$,
$ r_{pc}\simeq 2.7 \times 10^6 T_{dust}^{-2} L_{12}^{1/2}(FIR) \Omega^{-1/2}$.
Assuming T=50 K and guessing $\Omega / 4 \pi \simeq 0.2$, we get in this case, for the 
most luminous QUEST QSOs, a radius of about 1 kpc.

The critical challenge for the cool AGN-heated dust scenario
is the presence of strong PAH emission features in many of our QSOs and in the average
SEDs, which are so naturally associated with the FIR emission
if the latter is assumed to be due to star formation.
This is independent of  the assumed 
dimension of the cool dust region. However, the exact amount of starburst produced FIR emission 
requires more discussion.  In particular, 
the specific L(FIR)/L(PAH) ratio  observed in star forming ULIRGs 
and adopted here naturally leads to the
scenario of starburst dominated FIR emission discussed in the previous 
section. However, this ratio is known to depend on the conditions in the  ISM 
and can be, in some cases, significantly lower than the one assumed here.
This is known to be the case in quiescent 'cirrus-type' hosts.
In the extreme case of $\sim$10 times lower L(FIR)/L(PAH) ratio  
(known for some quiescent disks) a flat or a rising 
intrinsic AGN continuum out to FIR wavelengths remains possible. Cirrus 
type conditions cannot be excluded for the 
low PAH and FIR luminosity members of our 
sample which overlap with luminosities of quiescent galaxies (paper I), but 
become progressively unlikely at high star forming luminosities in the LIRG 
and ULIRG regime. More support for ULIRG-like FIR to PAH 
ratio in QSO hosts is given by the recent finding of a ULIRG-like ratio
in the $\sim$5 times more luminous Cloverleaf QSO \citep{lutz07}, where the observed L(PAH) 
is much higher than in known quiescent disks.

Considerations of cirrus emission add to the uncertainty on the $\lambda >30\mu$m slope of 
the intrinsic AGN SED derived in the previous section. They clearly leave 
the possibility of a flatter long wavelength slope for the QSOs with lower
L(PAH) and L(FIR). Most of these sources belong to our 
FIR-weak group with a global SEDs which is already falling gently at longer wavelengths. Thus, a dropping
FIR component must also be present in the
intrinsic AGN SED even if L(FIR)/L(PAH) is smaller than the one assumed here. As
for those sources with larger L(PAH) and 
L(FIR), the assumed ULIRG-like
ratio is favored because of the similar luminosity to the QUEST ULIRGs. 
Here again the intrinsic AGN SED must be falling at long wavelengths but with a considerable 
uncertainty on the slope.

To summarize, the
assumption of  AGN dominated FIR emission then requires a solution to both the PAH and 
dimension problems. It would also require an explanation for the almost
identical ratios between L(FIR) and L(\pahIR) observed in our QUEST
QSOs and in ULIRGs (paper I). We therefore consider the assumption of a starburst dominated
FIR emission to be more plausible for the
PG QSOs discussed here. For QSOs with a larger ratio
of AGN to host luminosity, the intrinsic AGN SED may 
dominate the emission  out to longer wavelengths.

\subsection{The AGN-starburst connection at low and high redshift}
The observations shown in Fig.~5 clearly suggest  significant correlations  between
the primary AGN continuum and the starburst produced emission (given our preferred explanation for
the origin of the FIR emission).
This was already pointed out in paper I where similar correlations relating e.g. L(6$\mu$m) and
L(60$\mu$m) were shown. This is of considerable theoretical interest since current galaxy and
BH evolution models do not predict a specific trend and only describe a general evolutionary relationship which
is not necessarily coeval. Such models, that are based on detailed numerical simulations (e.g. Hopkins et al.
2006; Volonteri et al. 2006; Granato et al. 2004) calculate the star formation rate and the BH
 growth rates as functions 
of cosmic time yet they definitely allow a time lag between the end of star formation and the 
commencement of black hole activity. Our observations of the QUEST sample do not require 
such a delay at redshifts 0.1--0.3. The sample may not represent all
AGNs but it gives a fair representation of unobscured sources at those redshifts.
This may indicate that enhanced SF activity, when present, 
includes the very central part of the galaxy. Thus, some of the starburst produced gas can find its way 
to the vicinity of the BH on a time scale which is short compared to the life time of the global, galactic  
scale starburst activity. More complicated scenarios including the obscured and unobscured phases of
a certain source (e.g. Hopkins et al. 2006) are also possible. 

A similar phenomenon may well be occurring  at redshifts much larger than 0.3.
Deep sub-mm and mm photometry has led to the detection of rest frame
submm and far-infrared dust emission from radio-quiet quasars
at redshifts up to 6.42 \citep[e,g,][]{omont01,isaak02,bertoldi03}.
Indirect arguments, like CO measurements, have been used to suggest  that this
emission is powered by star formation, implying that these quasars coexist with
extremely powerful $\gtrsim 10^{13}\Lsun$ starbursts. If the PAH to FIR
emission in these QSOs is similar to the one in ULIRGs, detection of PAH
on top of a strong continuum may be within the reach of \spitzer\ spectroscopy.
PAH emission from similar luminosity SMGs has been shown to be
detectable, e.g. \citet{lutz05}. Clearly, the extrapolation of the 
relationships found here to higher redshift, higher luminosity
AGNs depends on the ability to detect PAH features in such sources. There are already
some interesting upper limits in several extremely luminous QSOs (Maiolino et al. 2007)
and a  real detection in a high redshift high luminosity QSO with a strong
millimeter  flux (Lutz et al. 2007)

Having measured the L(5100)-L(FIR) relationship, we can estimate the relative growth rate of the central
BH and the galactic bulge, assuming all the observed star-forming activity contributes to the growth of the bulge.
For the QUEST QSOs, L(FIR)/L$_{\rm bol} \simeq 0.15$. Assuming 
a BH accretion efficiency $\eta$, 
we can convert the (primary) bolometric luminosity to BH mass growth rate
 and the observed FIR
luminosity to star formation rate.
This gives
$g$(bulge)/$g$(BH)$\sim 20 (\eta/0.1)$ where $g$ stands for growth
rate. This is more than 
an order of magnitude smaller than that required to explain the locally observed bulge and BH mass ratio
under the assumption of a similar duration for the two phenomena.
Thus, the numbers presented here may indicate that
the AGN activity phase is, on the average,
an order of magnitude shorter than the star-formation phase.
It would be interesting to carry out a similar analysis for higher redshift, higher luminosity AGNs.
For example, Steidel et al. (2002) suggest AGN fractions of about 3\% in $z\sim3$ UV-selected
samples, not very different from what was found here.

\section{Conclusions}

The main conclusion of the present work, and of paper I,  is that most and perhaps all of
the FIR luminosity of the QUEST QSOs is due to starburst activity.
This conclusion is based mostly on the tight correlation between the 
luminosity of the \pahIR\ feature and the FIR luminosity.
As explained in \S3, there are alternative relationships for L(FIR)/L(PAH)
that are different from the ones  used here but these we consider inapplicable to
the objects in question. While there are
clear \pahIR\ detections in only 11 of the sources studied here, most of 
the derived upper limits, as well as the composite spectrum of all sources 
{\it not showing} clear PAH emission, are consistent with this assumption 
(paper I). In this scenario, the starburst luminosity in our sample
is between $1.6 \times 10^{10}$ to $2.5\times 10^{12}$\Lsun\
corresponding to star formation rates of about 2 and 300 solar 
masses per year. The upper luminosity range is close to the luminosity of 
the most luminous starburst-dominated ULIRGs in our sample.

The assumption that most of FIR luminosity is due to starburst activity 
allows us to estimate the minimum temperature of the AGN heated dust in 
the central dusty structures of our QSOs. This temperature is of order 
200\,K. The inferred maximum dimensions of the torus in the simplest, 
continuous gas distribution case is of order 100$L_{12}^{1/2}$ pc, in 
reasonable agreement with all available observations.

Finally, our work shows that observed AGN SEDs can be misleading showing shapes  that are
quite different from the intrinsic SED.
We  suggest that future AGN models that try to reproduce the NIR-FIR spectrum of
type-I sources use the results  presented in Fig.~6 and Table~\ref{tab:seds} as
a more realistic representation of the intrinsic AGN spectrum. 

\begin{acknowledgements}

We thank Todd Boroson for kindly allowing us to use his optical data for 
PG quasars. We thank Amiel Sternberg for discussions and comments on the manuscript.
Funding for this work has been provided by the Israel Science 
Foundation grant 232/03. SV, DSR, and DCK were supported
in part by NASA contract 1263752 issued by JPL/Caltech.
HN acknowledges a Humboldt foundation prize and thanks the host 
institution, MPE Garching, where most of this work was performed.

\end{acknowledgements}

\end{document}